\documentstyle[12pt]{article}

\begin{document}

\begin{titlepage}
\begin{center}
        {\large \bf Dirac equation in some homogeneous space-times,\\
         separation of variables and exact solutions}

\vspace{1.0 cm}

V\'{\i}ctor M. Villalba\\
Centro de F\'{\i}sica,\\
            Instituto Venezolano de Investigaciones Cient\'{\i}ficas
(IVIC)\\
            Apartado 21827, Caracas $1020A$, Venezuela

\end{center}
\vspace{1.0 cm}

\begin{abstract}
\noindent In the present article, using a further generalization of the
algebraic method of separation of variables,
the Dirac equation is separated in a family of space-times where it
is not possible to find a complete set of first
order commuting differential operators. After separating variables,
the Dirac equation is reduced to a set of coupled
ordinary differential equations and some exact solutions corresponding
to cosmological backgrounds and gravitational waves
are computed in terms of hypergeometric functions. By passing, the
Klein Gordon equation in this background field is discussed.
\end{abstract}

\vspace{1.0 cm}
{\bf Pacs:} 11.10 Qr, 4.90. +e.
\medskip
\end{titlepage}

\section{Introduction}

\setcounter{equation}{0}

Since the appearance of the work of St\"ackel \cite{Stackel} a
considerable
amount of work has been devoted to the study of the problem of
separation of
variables for the Hamilton-Jacobi, Schr\"odinger and Klein-Gordon
equations,
in the search of exact solutions to different problems in mathematical
Physics. This problem is of particular importance in theoretical physics
because the construction of a quantum field theory in curved spaces and
the
definition of quantum vacuum\cite{Birrel} is impossible without a
detailed
study of the exact solutions of relativistic wave equations in curved
spaces. In this direction, and related to the problem of particle
creation
by strong gravitational fields,a great body of papers have been
published
during the last three decades after de appearence of the pioneers worls
by
Parker \cite{Parker} and \cite{Grib}. For a survey of the literature we
refer to \cite{Birrel,Grib1,Parker1}

Despite the remarkable success achieved in classifying symmetries and
separating variables for scalar wave equations, the corresponding
problem
for matrix partial differential equations, like the Dirac equation, a
system
of four coupled partial differential equations, is still far of being
completely solved. In this direction different approaches have been
proposed, most of then based on the experience acquired in studying
scalar
equations.

The interest in separating variables for the Dirac equation in curved
space
times goes back to the pioneer works of Taub \cite{Taub} and Podolsky
\cite
{Podolsky} At present there are different well established approaches to
tackle this problem, among them we can mention the St\"ackel spaces
method
by Bagrov {\it et al} \cite{Bagrov1,Bagrov2}, the Chandrasekhar's method
of
separation \cite{Chandr1,Chandr2}, the theory of factorizable systems by
Miller \cite{Miller1,Miller2}, and the symmetry operator approach \cite
{symm1,symm2}. Recently an algebraic method of separation of variables
\cite
{alg1,alg2,alg3} has been successfully applied in different
gravitational
and electromagnetic configurations. Regarding the question about the
equivalence among the different techniques of separation there are not
conclusive theorems showing that one, and some methods have shown to be
more
effective in the study of concrete gravitational configurations.
Recently,
Kalnins, Miller and Willians \cite{Miller3} have reviewed the use of
separation of variables in general relativity, and for instance, based
on
the idea of a factorizable system, they discuss the separation of
variables
for the Dirac equation in curved space-times. They exhibit a result
obtained
by Fels and Kamran \cite{symm2} showing that there exist separable
systems
for the Dirac equation in curved space-time that are not factorizable
and,
for instance, they exhibit a concrete example where this situation takes
place.

\begin{equation}
\label{1}ds^{2} = -dt^{2} + a^{2}(t)\left(dx^{2} + b^{2}(x)\left(dy^{2}
+
c^{2}(y)dz^{2}\right)\right)
\end{equation}
\noindent The metric (\ref{1}) is a St\"ackel space where the
Klein-Gordon
equation separates. It is worth noticing that the Robertson-Walker
universes
are particular cases of the line element (\ref{1}). It is the purpose of
the
present article to apply in an explicit way the algebraic method of
separation of variables and, after reducing the Dirac equation to a
system
of coupled ordinary differential equations to obtain exact solutions of
the
Dirac equation in some gravitational homogeneous backgrounds.

First, we proceed to separate variables for the Klein-Gordon equation in
the
metric (\ref{1}). The relativistic equation for a spinless scalar
particle
in a curved background reads

\begin{equation}
\label{2}%
\pmatrix{g^{\alpha \beta }\partial_{\alpha }\partial_{\beta
}~-~g^{\alpha \beta }\Gamma ^{\lambda }_{\alpha \beta }\partial
_{\lambda }~-~\xi R~-~m^{2}}
\Phi = 0
\end{equation}
\noindent where $\xi $ is a dimensionless constant and $R$ is the
constant
of curvature. The specialization of equation (\ref{2}) to the metric
(\ref{1}%
) gives

\begin{eqnarray}
\label{3}
(-\partial^{2}_{t}- {3\over a}a,_{t}\partial _{t} + {1\over
a^{2}}\pmatrix{\partial ^{2}_{x}+{2\over b} b,_{x}\partial _{x}} +
{1\over a^{2}b^{2}}\pmatrix{\partial ^{2}_{y}&+&{1\over c}
c,_{y}\partial _{y}}\nonumber\\
+{1\over a^{2}b^{2}c^{2}} \partial ^{2}_{z} - m^{2})\Phi=0
\end{eqnarray}

\noindent where we have equated to zero the coupling constant $\xi $.
Substituting into (\ref{3}) a solution of the form

\begin{equation}
\label{4}\Phi = T(t)X(x)Y(y)Z(z)
\end{equation}
\noindent we find that Eq. (\ref{3}) separates as follows

\begin{equation}
\label{5}\left[\partial ^{2}_{t} + {\frac{3}{a}}a,_{t}\partial _{t} +
m^{2}
+ {\frac{\lambda ^{2}}{a^{2}}}\right]T(t) = 0
\end{equation}
\begin{equation}
\label{6}\left[\partial ^{2}_{x} + {\frac{2}{b}} b,_{x}\partial _{x} +
\lambda ^{2} - {\frac{\sigma ^{2}}{b^{2}}}\right]X(x) = 0
\end{equation}
\begin{equation}
\label{7}\left[\partial^{2}_{y} + {\frac{1}{c}} c,_{y}\partial _{y} +
\sigma
^{2} - {\frac{\delta ^{2}}{c^{2}}}\right]Y(y) = 0
\end{equation}
\begin{equation}
\label{8}\left[\matrix{\partial ^{2}_{z}&+&\delta ^{2}}\right]Z(z) = 0
\end{equation}
\noindent where a comma indicates derivation on the corresponding
variable,
and $\lambda , \sigma $, and $\delta $ are constants of separation; then
the
Klein-Gordon equation is completely separable in the background field
(\ref
{1}). The present article is structured as follows, in section 2,
following
the consecutive and pairwise schemes of the algebraic method of
separation
of variables for the Dirac equation in the background field (\ref{1}) ,
we
reduce the Dirac equation to a set of coupled ordinary differential
equations. In section 3, applying the results obtained in Sec. 2 we
solve
the Dirac equation in some homogeneous universes.

\section{Separation of variables}

\setcounter{equation}{0} \noindent Now we proceed to write the covariant
generalization of the Dirac equation in the metric (\ref{1})

\begin{equation}
\label{9}\left\{{\tilde{\gamma}}^{\mu }(\partial _{\mu } - \Gamma _{\mu
}) +
m\right\}\tilde{\Psi} ~=~0
\end{equation}
\noindent where $\tilde{\gamma }^{\mu }$ are the curved gamma matrices
satisfying the relation,

\begin{equation}
\label{10}\left\{{\tilde{\gamma}}^{\mu }, \tilde{\gamma}^{\nu
}\right\}_{+}=
2g^{\mu \nu }
\end{equation}
\noindent and $\Gamma _{\mu }$ are the spin connections \cite{Brill}.
\begin{equation}
\label{11}\Gamma _{\lambda } = {\frac{1}{4}} g_{\mu \alpha } [(\partial
b^{\beta }_{\nu } /\partial x^{\lambda }) a^{\alpha }_{\beta } - \Gamma
^{\alpha }_{\nu \lambda }] s^{\mu \nu }
\end{equation}
where

\begin{equation}
\label{12}s^{\mu \nu } = {\frac{1}{2}}(\tilde\gamma ^{\mu } \tilde\gamma
^{\nu } -\tilde\gamma ^{\nu } \tilde\gamma ^{\mu })
\end{equation}
\noindent and the matrices $b^{\beta }_{\alpha }$ , $a^{\alpha }_{\beta
}$
establish the connection between the Dirac matrices $(\tilde\gamma )$ on
a
curved spacetime and the Minkowski space $(\gamma)$ Dirac matrices as
follows:

\begin{equation}
\label{13}\tilde\gamma _{\mu } = b^{\alpha }_{\mu }{\gamma }_{\alpha
}\qquad
\tilde\gamma ^{\mu } = a^{\mu }_{\beta }{\gamma}^{\beta }
\end{equation}
Choosing to work in the diagonal tetrad gauge for $\tilde{\gamma }^{\mu
}$,

\begin{equation}
\label{14}\tilde{\gamma }^{0}={\gamma}^{0}, \hspace{.3cm}
\tilde{\gamma}%
^{1}=a^{-1}\gamma ^{1}, \hspace{.3cm}
\tilde{\gamma}^{2}=a^{-1}b^{-1}\gamma
^{2}, \hspace{.3cm} \tilde{\gamma }^{3}=a^{-1}b^{-1}c^{-1}\gamma^{3}
\end{equation}
\noindent where $\gamma ^{\mu }$ are the standard Dirac flat matrices.
Then,
we have that the Dirac equation (\ref{9}) takes the form

\begin{equation}
\label{15}\left\{\gamma ^{0}\partial _{t} + {\frac{1}{a}} \gamma
^{1}\partial _{x} + {\frac{1}{ab}} \gamma ^{2}\partial _{y} +
{\frac{1}{abc}}
\gamma ^{3}\partial _{z} + m\right\}\Psi = 0
\end{equation}
\noindent where we have introduce the spinor $\Psi $,

\begin{equation}
\label{16}\tilde{\Psi } = a^{-3/2}b^{-1}c^{-1/2}\Psi
\end{equation}
\noindent Applying the algebraic method of separation of variables
[12-14]
it is possible to write eq. (\ref{15}) as a sum of two first order
differential operators $\hat{K}_{1}$ , $\hat{K}_{2}$ satisfying the
ra,X:ion

\begin{equation}
\label{17}\left[\hat{K}_{1},\hat{K}_{2}\right]= 0, \hspace{2cm}
\left\{\hat{K%
}_{1}+ {\hat{K}}_{2}\right\}\Phi = 0
\end{equation}
\begin{equation}
\label{18}{\hat{K}}_{1}\Phi = \lambda \Phi = -{\hat{K}}_{2}\Phi
\end{equation}
\noindent where

\begin{equation}
\label{19}\Psi = \gamma ^{1}\gamma ^{2}\gamma ^{3}\Phi ,
\end{equation}
\begin{equation}
\label{20}\hat{K}_{1} = a\left(\gamma ^{0}\partial _{t}+m\right)\gamma
^{1}\gamma ^{2}\gamma ^{3},
\end{equation}
\begin{equation}
\label{21}\hat{K}_{2} = \left(\gamma ^{1}\partial _{x}
+{\frac{1}{b}}\gamma
^{2}\partial _{y} + {\frac{1}{bc}}\gamma ^{3}\partial _{z}\right)\gamma
^{1}\gamma ^{2}\gamma ^{3}
\end{equation}
\noindent in this way, we have separated the time variable $t$, from the
spatial $x, y$ and $z$ variables. The problems arises when we try to
reduce
equation (\ref{21}) to the form (\ref{17}). It is not difficult to see
that
there are not separating matrices allowing that step. In order to go
further
we write equation $(\hat{K}_{2}~+~\lambda )\Phi ~=~0$ as follows:

\begin{equation}
\label{22}\pmatrix{\hat{K}_{3}+\hat{K}_{4}\gamma ^{1}\gamma ^{2}}\Phi
{}~=~0
\end{equation}
\noindent where $\hat{K}_{3}$ and $\hat{K}_{4}$ are two commuting
differential operators given by the expressions

\begin{equation}
\label{23}\hat{K}_{3}~=~b%
\pmatrix{\gamma ^{2}\gamma ^{3}\partial _{x}~+~\lambda }, \hspace{2cm}
\hat{K%
}_{4}~=~\left(\gamma ^{2}\gamma ^{3}\partial
_{y}~+~{\frac{1}{c}}\partial
_{3}\right),
\end{equation}
\noindent After introducing the auxiliary spinor $\Sigma $

\begin{equation}
\label{24}\Phi ~=~\pmatrix{\hat{K}_{4}~+~\gamma ^{1}\gamma
^{2}\hat{K}_{3}}%
\Sigma
\end{equation}
\noindent we reduce equation (\ref{22}) to the form;

\begin{equation}
\label{25}\left\{N_{1} + \hat{N}_{2}\right\}\Sigma ~=~0, \hspace{1cm}
\left[%
\matrix{\hat{N}_{1},&\hat{N}_{2}}\right]~=~0
\end{equation}
\noindent with

\begin{equation}
\label{26}\hat{N}_{2}\Sigma = - \hat{N}_{1}\Sigma = k^{2}\Sigma
\end{equation}
\noindent where

\begin{equation}
\label{27}\hat N_1~=~%
\pmatrix{b^{2}\partial ^{2}_{x}+b\partial _{x}b\partial _{x}-\lambda
\gamma ^{2}\gamma ^{3}\partial _{x}b+\lambda ^{2}b^{2}}%
,
\end{equation}
\begin{equation}
\label{28}\hat N_2~=~%
\pmatrix{\partial ^{2}_{y}+i\gamma ^{2}\gamma ^{3}k_{z}c^{-2}\partial
_{y}c-k^{2}_{z}c^{-2}}
\end{equation}
\noindent $k_z$ in (\ref{28}) is the eigenvalue of the operator
$-i\partial
_z$.

\noindent Choosing to work in the following representation of the Dirac
matrices:

\begin{equation}
\label{29}\gamma ^{0}~=~\pmatrix{-i&0\cr 0&i}, \gamma ^{1}~=~%
\pmatrix{0 &\sigma ^{3}\cr \sigma ^{3} &0}, \gamma ^{2}~=~%
\pmatrix{0 &\sigma ^{2}\cr \sigma ^{2} &0}, \gamma ^{3}~=~%
\pmatrix{0 &\sigma ^{1}\cr \sigma ^{1} &0},
\end{equation}
\medskip
\noindent we have that the spinor $\Sigma $ takes $te$ form

\begin{equation}
\label{30}\Sigma ~=~\pmatrix{p(x)A(y)\cr q(x)B(y)\cr r(x)C(y)\cr
s(x)D(y)}
\end{equation}
\noindent the functions $a(x), b(x), c(x)$ and $d(x)$ satisfy the
differential equation

\begin{equation}
\label{31}\pmatrix{b(d_{x}&+&\mp i\lambda )}\pmatrix{b(d_{x}+&\pm
i\lambda )}%
X(x) + k^{2}X(x)~=~0
\end{equation}
\noindent where the upper signs correspond to $p(x), r(x)$, and the
lower
signs are related to $q(x)$ and $s(x)$.

\noindent The functions $A(y), B(y), C(y), D(y)$ are solution of the
equation

\begin{equation}
\label{32}\pmatrix{d_{y}&\pm &k_{z}c^{-1}}\pmatrix{d_{y}&\mp
&k_{z}c^{-1}}%
Y(y) - k^{2}Y(y)~=~0
\end{equation}
\noindent where the upper signs correspond to A(y) and $C(y)$, and the
lower
ones correspond to $B(y)$ and $D(y)$. After substituting the expression
(\ref
{30}) into (\ref{24}) we arrive at

\begin{equation}
\label{33}\Phi ~=~ik%
\pmatrix{p(x)B(y)(\varepsilon-\omega^{-1})\alpha\cr
q(x)A(y)(\varepsilon+\omega^{-1})\alpha\cr
r(x)D(y)(\varepsilon-\omega^{-1})\beta\cr
s(x)C(y)(\varepsilon+\omega^{-1})\beta\cr}%
exp(ik_{z}z)
\end{equation}
\noindent where $\alpha $ and $\beta $ are functions depending on the
variable t. The functions $A, B, C, D$, and $p, q, r, s$ and are related
by
the following systems of coupled partial equations:

\begin{equation}
\label{34}(d_{y} - {\frac{k_{z}}{c}})\pmatrix{A\cr C}=-k\epsilon
\pmatrix{B\cr D}, (d _{y}+ {\frac{k_{z}}{c}})\pmatrix{B\cr D}=k/\epsilon

\pmatrix{A\cr C}
\end{equation}
\begin{equation}
\label{35}b(d_{x} + i\lambda ) \pmatrix{p\cr r}~=~-ik\omega
\pmatrix{q\cr s}
, \\ b(d_{x} - i\lambda )\pmatrix{q\cr s} ~=~-ik/\omega \pmatrix{p\cr r}

\end{equation}
\noindent where $\epsilon $ and $\omega $ are arbitrary constants. From
(\ref
{33}), and (\ref{20}) we have, $p(x)~=~r(x)$, $q(x)~=~s(x)$,
$A(y)~=~C(y)$,
and $B(y)~=~D(y)$. The functions $\alpha $ and $\beta $ satisfy the
system:

\begin{equation}
\label{36}-a(\partial _{t} + {\rm im})\beta ~=~\lambda \alpha ,\qquad
a(\partial _{t} - {\rm im})\alpha ~=~\lambda ~\beta
\end{equation}
\noindent Finally, from (\ref{16}), (\ref{19}) and (\ref{20}) we find
that
the solution of the Dirac equation (\ref{9}) reads

\begin{equation}
\label{37}\tilde{\Psi} = a^{-3/2}b^{-1}c^{-1/2}%
\pmatrix{p(x)B(y)(\varepsilon-\omega^{-1})\beta(t)\cr
q(x)A(y)(\varepsilon+\omega^{-1})\beta(t)\cr
p(x)B(y)(\varepsilon-\omega^{-1})\alpha(t)\cr
q(x)A(y)(\varepsilon+\omega^{-1})\alpha(t)}
exp(ik_{z}z)
\end{equation}
Now, we proceed separate variables for the Dirac equation (\ref{11})
using a
pairwise scheme. Eq. (\ref{15}) can be reduced to the form
(\ref{17})-(\ref
{18})

\begin{equation}
\label{38}\hat{K}_{1}~=~-iab\left[\gamma ^{0}\partial _{t} +
{\frac{\gamma
^{1}}{a}}\partial _{x} + m\right]\gamma ^{1}\gamma ^{0}
\end{equation}
\begin{equation}
\label{39}\hat{K}_{2}~=~-i\left[\gamma ^{2}\partial _{y} + {\frac{\gamma
^{3} }{c}}\partial _{z}\right]\gamma ^{1}\gamma ^{0}
\end{equation}
\begin{equation}
\label{40}\Phi = \gamma ^{1}\gamma ^{0}\Psi
\end{equation}
\noindent where, in this case, the commuting differential operators
$\hat{K}%
_{1}$ and $\hat{K}_{2}$ depend on two spacetime variables. the
expression $
\hat{K}_{2}\Phi ~+~k \Phi ~=~0$ can be separated as follows

\begin{equation}
\label{41}\hat{K}_{3}~=~c\left[%
\matrix{\gamma ^{2}\gamma ^{3}\partial _{y}+ik\gamma ^{3}\gamma
^{1}\gamma ^{0}}%
\right], \hspace{.5cm} \hat{K}_{4}~=~\partial _{z}
\end{equation}
\noindent where

\begin{equation}
\label{42}\left[ \matrix{\hat{K}_{3},&\hat{K}_{4}}\right]
{}~=~0,\hspace{.5cm}%
\pmatrix{\hat{K}_{3}&+&\hat{K}_{4}}\Omega =0,\hspace{.5cm}\Omega
{}~=~\gamma
^3\gamma ^1\gamma ^0\Phi
\end{equation}
\noindent The separation of variables in the equation $\hat K_1\Phi
{}~=~k\Phi
$ cannot be achieved in terms of first order commuting differential
operators, it is advisable to rewrite this equations as follows:

\begin{equation}
\label{43}\left\{ \left[ \gamma ^0\partial _x+i{\frac kb}\right]
+a\left[
\gamma ^0\partial _t+m\right] \gamma ^1\gamma ^0\right\} \Phi =\left\{
\hat
L_1+\hat L_2\gamma ^1\gamma ^0\right\} \Phi =0
\end{equation}
\noindent it is easy to see that the operators $\hat L_1$ and $\hat L_2$
commute,

\noindent Let us introduce the auxiliary function $\eta $ defined by,
\begin{equation}
\label{44}\Phi ~=~\left(a\left[\gamma ^{0}\partial _{t}+m\right]\gamma
^{1}\gamma ^{0}+\left[\gamma ^{0}\partial_{x}-i{\frac{k}{b}}%
\right]\right)\eta
\end{equation}

\noindent Eq. (\ref{40}) reduces to

\begin{equation}
\label{45}\left\{N_{1} + \hat{N}_{2}\right\}\eta = 0\qquad \left[%
\matrix{\hat{N}_{1},&\hat{N}_{2}}\right] = 0
\end{equation}
\noindent where

\begin{equation}
\label{46}\hat{N}_{2}\eta = - \hat{N}_{1}\eta = \lambda ^{2}\eta
\end{equation}
\noindent and the operators $\hat{N}_{1}$ and $\hat{N}_{2}$ read:

\begin{equation}
\label{47}\hat{N}_{1}~=~\left[a\left(\gamma ^{0}\partial
_{t}+m\right)\right]\left[a\left(-\gamma ^{0}\partial
_{t}+m\right)\right]
\end{equation}
\begin{equation}
\label{48}\hat{N}_{2}~=~\left[\gamma ^{0}\partial _{x}+i{\frac{k}{b}}%
\right]\left[\gamma ^{0}\partial _{x}-i{\frac{k}{b}}\right]
\end{equation}
\noindent using the representation (\ref{29}) for $\gamma ^{0}$ and
(\ref{46}%
)-(\ref{47}) we can write the spinor $\eta $ as follows

\begin{equation}
\label{49}\eta ~=~\pmatrix{u(t)U(x)\cr v(t)V(x)\cr w(t)W(x)\cr z(t)Z(x)}

\end{equation}
\noindent applying (\ref{44}) to (\ref{49}), we have that $\Phi $ takes
the
form:

\begin{equation}
\label{50}\Phi ~=~%
\pmatrix{(-\delta^{-1}+\sigma)u(t)W(x)f(y)\cr
(+\delta^{-1}+\sigma)u(t)W(x)g(y)\cr (+\delta
-\sigma^{-1})w(t)U(x)f(y)\cr (-\delta -\sigma^{-1})w(t)U(x)g(y)}
exp (ik_{z}z)
\end{equation}
\noindent where the functions $u(t), w(t), U(x)$ and $W(x)$ satisfy the
coupled systems of equations

\begin{equation}
\label{51}(d_{x} - {\frac{k}{b}})W~= i\lambda /\sigma U, (d_{x} +
{\frac{k}{b%
}})U~=i\lambda \sigma W
\end{equation}
\begin{equation}
\label{52}a(d_{t} - {\rm im})u~= \lambda \delta w, a(d_{t} + {\rm
im})w~=
-\lambda /\delta u
\end{equation}
\noindent $\delta $ and $\sigma $ are arbitrary constants. The functions
$%
f(y)$ and $g(y)$ can be determined by means of the equations (\ref{39})
and (%
\ref{41}) giving as result

\begin{equation}
\label{53}(-\delta ^{-1}+ \sigma )(\partial
_{y}-{\frac{k_{z}}{c}})f(y)~=~ik
(\delta ^{-1}+ \sigma )g(y)
\end{equation}
\begin{equation}
\label{54}(\delta ^{-1}+ \sigma )(\partial _{y} + {\frac{k_{z}}{c}}%
)g(y)~=~ik (-\delta ^{-1}+ \sigma )f(y)
\end{equation}
\noindent then, from (\ref{16}) and (\ref{40}) we find that, the spinor
$
\tilde{\Psi }$ reads

\begin{equation}
\label{55}\tilde \Psi =a^{-3/2}b^{-1}c^{-1/2}%
\pmatrix{(\delta-\sigma^{-1})w(t)U(x)f(y)\cr
(\delta+\sigma^{-1})w(t)U(x)g(y)\cr (\delta^{-1}-\sigma)u(t)W(x)f(y)\cr
(\delta^{-1}+\sigma)u(t)W(x)g(y)}%
exp(ik_zz)
\end{equation}
\noindent here some remarks are in order, using the algebraic method of
separation of variables we have reduced the problem of solving the Dirac
equation (\ref{9}), in the curved background (\ref{1}), to finding
solutions
of the coupled systems of equations (\ref{34})-(\ref{36}) and
(\ref{51})-(%
\ref{54}), for the consecutive and pairwise schemes of separation
respectively. Since $\hat K_3$ in (\ref{23}), and $\hat N_1$ in
(\ref{27}),
commute with $mI~+~n\gamma ^2\gamma ^3$ where $m$ and $n$ are arbitrary
constants, the expression $(mI~+~n\gamma ^2\gamma ^3)\Phi $ is also
solution
of (\ref{26}). In this way, we have that the set of solutions
(\ref{34})-(%
\ref{36}) is equivalent to those obtained from (\ref{51})-(\ref{54})

\section{Exact solutions}

\setcounter{equation}{0}

The number of gravitational space-times where the Dirac equation is
completely separable iqdMQI%
Q"=
b
9X
5=9"!5A
Q
$X9e interested in studying solutions of the
Einstein field equations with cosmological constant $\Lambda$ f and the
form
studied in the preceding section, we have as concrete examples the de
Sitter
and anti de-Sitter universes.

The physical importance of these universes has been widely discussed in
the
literature and the computations of exact solutions of scalar and spin
1/2
particles is of help in understanding quantum effects in these
backgrounds
\cite{shishkin,Galtsov1,Galtsov2,Otchik,Gibbons}. Here in this section
we
compute exact solutions of the Dirac equation in different charts where
the
metric is expressed in the comoving frame. Now we proceed to write the
Dirac
equation in an expanding cosmological Robertson-Walker space-time
expressed
in the comoving frame, where the conformal factor $e^{\alpha }$ depends
only
on the time parameter given by $\eta$

\begin{equation}
\label{RW}ds^{2} = e^{\alpha (\eta )}\{dr^{2} + \xi ^{2}(r) d \Omega
^{2} -
d \eta ^{2}\}
\end{equation}
\noindent and the radial factor $\xi $ takes the values

\medskip
\begin{equation}
\xi (r) = \left\{
\begin{array}{cc}
r &  \\
\sinh(r) &  \\
\sin(r) &
\end{array}
\right.
\end{equation}
\medskip
\noindent for spatially flat, open and closed universes respectively.

\noindent Choosing to work in the diagonal tetrad $a^{\alpha }_{\beta }$

\medskip
\begin{equation}
a^{\alpha }_{\beta } = diag%
\pmatrix{e^{-\alpha /2},e^{-\alpha /2},e^{-\alpha /2}\xi
(r)^{-1},e^{-\alpha /2}(\xi (r)\sin \theta )^{-1}}
\end{equation}

\noindent we obtain that the Dirac equation in the Robertson-Walker
background field (\ref{RW}) reads

\begin{equation}
\label{DRW}\left( \frac{{\gamma }^0}{e^{\alpha /2}}\partial _\eta
+\frac{{%
\gamma }^1}{e^{\alpha /2}}\partial _r+\frac{{\gamma }^2}{e^{\alpha
/2}\xi (r)%
}\partial _\vartheta +\frac{{\gamma }^3}{e^{\alpha /2}\xi (r)\sin \theta
}%
\partial _\varphi +m\right) \Psi =0
\end{equation}

\medskip
where $\Psi $ is related to $\tilde \Psi $ by

\begin{equation}
\tilde \Psi =\xi (r)^{-1}(\sin \theta )^{-1/2}e^{-3\alpha /4}\Psi
\end{equation}
Eq. ({\ref{DRW}) can be written in the form (\ref{17}) with,
\begin{equation}
{\hat K}_1=-i\{{\gamma }^2\partial _\theta +\frac{{\gamma }^3}{\sin
\theta }%
\partial _\varphi \}{\gamma }^1{\gamma }^0
\end{equation}
\begin{equation}
{\hat K}_2=-i\xi (r)\{({\gamma }^0\partial _0+{\gamma }^1\partial
_1)+me^{\alpha /2}\}{\gamma }^1{\gamma }^0
\end{equation}
\begin{equation}
\Phi ={\gamma }^1{\gamma }^0\Psi
\end{equation}
\noindent The operator ${\hat K}_1$ corresponds to the ''momentum''
obtained
by Brill and Wheeler \cite{Brill} in the problem of separation of
variables
of Dirac equation in the Schwarzschild metric. In order to obtain the
physical angular momentum ${\bf K}$, it is necessary to apply the
unitary
transformation ${\bf S}$ relating Dirac matrices in the diagonal tetrad
gauge to the Dirac matrices in the cartesian gauge: }

\begin{equation}
{\bf S \gamma }^{\nu } {\bf S}^{-1} = h^{\nu }_{\beta } \tilde{\gamma }%
^{\beta }= \gamma ^{\nu }_{c}
\end{equation}
\begin{equation}
{\bf S }\Psi = \Psi _{c}
\end{equation}
\noindent where $\gamma ^{\nu }_{c}$ are Dirac matrices in the cartesian
tetrad gauge. The matrix transformation ${\bf S}$ reads:

\begin{equation}
{\bf S}=\exp (-{\frac \varphi 2}{\gamma }^1{\gamma }^2)\exp (-{\frac
\theta 2%
}{\gamma }^3{\gamma }^1){\cal N}{}
\end{equation}

\noindent where
\begin{equation}
{\cal N}={\frac 12}({\gamma }^1{\gamma }^2+{\gamma }^2{\gamma
}^3+{\gamma }^3%
{\gamma }^1+1)
\end{equation}

\noindent the system of ordinary differential equations associated with
the
time dependence of the line element (\ref{RW}) can be written as
follows.

\begin{equation}
\label{first}\left({\frac{d}{d\eta}}-ime^{\alpha /2}\right)u = -i\lambda
v
\end{equation}
\begin{equation}
\label{second}\left({\frac{d}{d\eta}}+ime^{\alpha /2}\right)v =
-i\lambda u
\end{equation}
Eqs. (\ref{first}), (\ref{second}) can be obtained from (\ref{52}) by
putting $\delta=-i$ and redefining the time parameter. Since the
equations
for the radial dependence $\xi(r)$ were already solved in Ref. \cite
{villalba1}, we proceed to compute the exact solutions of the system
(\ref
{first}) and (\ref{second}) in the de-Sitter and anti de-Sitter
universes
expressed in the comoving frame.

The de Sitter line element expressed in the static chart reads

\begin{equation}
ds^{2} = -\pmatrix{1-{1\over 3}\Lambda r^{2}}dT^{2}+
\pmatrix{1-{1\over 3}\Lambda r^{2}}^{-1}dr^{2}+ r^{2}d\Omega ^{2}
\end{equation}
\noindent The above expression can be reduced to the form (\ref{RW})
with
spatially closed curvature

\begin{equation}
\label{carta1}ds^{2} = {\frac{3}{\Lambda }}{\csc}^{2}\eta
\pmatrix{-d\eta ^{2}+d\kappa ^{2}+\sin ^{2}\kappa d\Omega ^{2}}
\end{equation}
\noindent by means of the transformation

\begin{equation}
\pmatrix{{\Lambda \over 3}}^{1/2}T =\hbox{ arctanh }(\cos \kappa
\hbox{ cosec}\eta )
\end{equation}
\begin{equation}
\pmatrix{{\Lambda \over 3}}^{1/2}r =\hbox{ cosec}\eta \sin \kappa
\end{equation}
\noindent notice that the chart (\ref{carta1}) covers all the manifold.

\noindent Substituting the value of $e^{\alpha/2}= \pmatrix{{3\over
\Lambda}}%
^{1/2} \frac{1}{\sin \eta}$ into (\ref{first})-(\ref{second}) we obtain

\begin{equation}
\label{uno}\left( {\frac d{d\eta }}-im{(\frac 3\Lambda )}^{1/2}{\frac
1{\sin
\eta }}\right) u=-i\lambda w
\end{equation}
\begin{equation}
\label{dos}\left( {\frac d{d\eta }}+im{(\frac 3\Lambda )}^{1/2}{\frac
1{\sin
\eta }}\right) w=-i\lambda u
\end{equation}
The system of equations (\ref{uno})-(\ref{dos}) has as solution
\begin{equation}
w(\eta )=c_0\sin {}^p(\eta )\sin ({\frac \eta 2})F({\frac 12}-\lambda
+p,{%
\frac 12}+\lambda +p,{\frac 32}+p,{\frac{1-\cos \eta }2})
\end{equation}
\begin{equation}
u(\eta )=ic_0{\frac{{\frac 12}+p}\lambda }\sin {}^p(\eta )\cos ({\frac
\eta 2%
})F({\frac 12}-\lambda +p,{\frac 12}+\lambda +p,{\frac
12}+p,{\frac{1-\cos
\eta }2})
\end{equation}
where $F(a,b,c,z)$ are the Gauss hypergeometric functions, and $p$ is
\begin{equation}
p={\rm im}\pmatrix{{3\over \Lambda }}^{1/2}
\end{equation}
\noindent Another chart where it is possible to write the $dS$ universe
with
time dependent conformal factor is

\begin{equation}
ds^{2} = {\frac{3}{\Lambda }} \eta ^{-2}%
\pmatrix{-d\eta ^{2}+d\kappa ^{2}+\kappa ^{2}d\Omega ^{2}}
\end{equation}
\noindent here the metric is spatially flat. The transformation relating
the
static coordinates $T, r$ with $\eta $ and $\kappa $ are

\begin{equation}
\pmatrix{{\Lambda \over 3}}^{1/2}r = {\frac{\kappa }{\eta }}
\end{equation}
\begin{equation}
\pmatrix{{\Lambda \over 3}}^{1/2}T =\hbox{ arctanh }%
\pmatrix{{\kappa \over \eta }}
\end{equation}
\noindent In the present case we have that $e^{\alpha/2 }=
\pmatrix{{3\over \Lambda }}^{1/2}{\frac{1}{\eta }}$ , substituting this
expression into the system of equations (\ref{first})-(\ref{second}), we
have

\begin{equation}
\left({\frac{d}{d\eta }}+im{(\frac{3}{\Lambda})}^{1/2}{\frac{1}{\eta }}%
\right)w = -i\lambda u
\end{equation}
\begin{equation}
\left({\frac{d}{d\eta }}-im{(\frac{3}{\Lambda})}^{1/2}{\frac{1}{\eta }}%
\right)u = -i\lambda w
\end{equation}
\medskip
\noindent The solution of this system is

\begin{equation}
w(\eta ) = a_{0} \eta ^{1/2} Z_{{\rm im} + 1/2} (\lambda \eta )
\end{equation}
\begin{equation}
u(\eta ) = ia_{0}\eta ^{1/2} Z_{{\rm im} - 1/2} (\lambda \eta )
\end{equation}
\noindent where $a_{0}$ is an arbitrary constant, and $Z_{\alpha}$ is
the
general solution of the Bessel equation.

The anti de-Sitter universe can be written in the static coordinates $T$
and
$r$ as follows

\begin{equation}
\label{adS}ds^{2} = -\pmatrix{1+{1\over 3}\Lambda r^{2}}dT^{2}+
\pmatrix{1+{1\over 3}\Lambda r^{2}}^{-1}r^{2}+ r^{2}d\Omega^{2}
\end{equation}
\noindent this cosmological universe also can be described in the
comoving
frame coordinates $\eta $ and $\kappa $ with spatially open curvature:

\begin{equation}
\label{chart2}ds^{2} = {\frac{3}{\Lambda }}\hbox{ sech}^{2}\eta
\pmatrix{-d\eta ^{2}+d\kappa ^{2}+\sinh ^{2}\kappa d\Omega ^{2}}
\end{equation}
\noindent the transformation of coordinates relating both charts
(\ref{adS})
and (\ref{chart2}) is

\begin{equation}
\pmatrix{{\Lambda \over 3}}^{1/2}T = -\arctan (\cosh \kappa \hbox{
cosech}%
\eta )
\end{equation}
\begin{equation}
\pmatrix{{\Lambda \over 3}}^{1/2}r =\hbox{ sech}\eta \sinh \kappa
\end{equation}
\noindent here the conformal factor $e^{\alpha/2}$ is

\begin{equation}
e^{\alpha/2} = \pmatrix{{3\over \Lambda }}^{1/2}{\frac{1}{\cosh \eta }}
\end{equation}
\noindent substituting the above expression into
(\ref{first})-(\ref{second}%
) we obtain:

\begin{equation}
w(\eta ) = a_{0}\sinh ^{p}(\eta )%
\pmatrix{\cosh ({\eta \over 2})+\hbox{isinh}({\eta \over 2})} F
({\frac{1}{2}%
} - i\lambda + k, {\frac{1}{2}} + i\lambda + k, {\frac{3}{2}} + k,
\frac{%
1+i\sinh \eta}{2})
\end{equation}
\begin{eqnarray}
u(\eta ) = ia_{0} {{1\over 2} + k\over \lambda } \sinh ^{k}(\eta )
\pmatrix{\cosh ({\eta \over 2})-i\sinh(\frac{\eta}{2})} \\ \nonumber
x F({1\over 2} - i\lambda  + k,{1\over 2} + i\lambda  + k,{1\over 2} +
k, \frac{1+i\sinh \eta}{2})
\end{eqnarray}
In this way, we have obtained exact solutions of the Dirac equation in
the
de-Sitter and anti de-Sitter Universes expressed in the comoving frame
coordinates. These examples do not exhaust all the possible homogeneous
universes where it is possible to express the exact solutions of the
Dirac
equation in terms of special functions, namely we have different
configurations with matter and radiation where the above one is
possible.

Finally, we show an example of a non diagonal metric where the
separation of
variables in the Dirac equation is possible as well as the computation
of
their exact solutions:
\begin{equation}
\label{Bertotti}ds^{2} = -du dv + b^{2}(u)\left[%
\pmatrix{dy^{2}+c^{2}(y)dz^{2}}\right]
\end{equation}
\noindent the case where $c(y)~=~1$ can be regarded as a degenerate $pp$
wave \cite{Kramer}, for $c(y)~=~y$ we also have a similar situation.
First,
we proceed to separate the spatial variables $y$ and $z$ from $u$ and
$v$,
and, since we have to deal with singular Dirac matrices, we do not
follow
the scheme proposed for the diagonal metric (\ref{1}). The Dirac
equation in
the background field (\ref{Bertotti}) reads
\begin{equation}
\left\{\gamma^{u}\partial_{u}+ \gamma^{v}\partial_{v}+m +
\frac{\gamma^{2}}{b%
}\partial_{2} + \frX
\end{equation}
after separating the space variables from $u$ and $v$ we have,
\begin{equation}
\label{eq}\left\{\left(\left[\gamma^{u}\partial_{u} +
\gamma^{v}\partial_{v}\right) + m\right]\gamma^{0}\gamma^{1}
-i\frac{K}{b}%
\right\}\Phi = 0
\end{equation}
where $K$ is the constant of separation associated with the operator
$\hat{K}%
_{2}$ given by the equation (\ref{31}). and,
\begin{equation}
u=t-x, \qquad v=t+x, \qquad \gamma^{u}= \gamma^{0}- \gamma^{1}, \qquad
\gamma^{v}= \gamma^{0}+ \gamma^{1},
\end{equation}
the solution of eq.(\ref{eq}) in the representation for the gamma
matrices (%
\ref{29}) can be written as follows
\begin{equation}
\Phi=\pmatrix{-(m+2p_{v}-iK/b)\eta\cr i\sigma^{3}(m-2p_{v}-iK/b)\eta
\cr}q,
\end{equation}
where $\eta$ is a two component spinor depending on the variables $y$
and $z$%
, and $q$ is a function satisfying the equation
\begin{equation}
4ip_{v}\partial_{u}q = -(-K^{2}/b^{2} +m^{2})q
\end{equation}
which has as solution,
\begin{equation}
q=\exp(\frac{i}{4p_{v}}\int(K^{2}/b^{2}+m^{2})du)
\end{equation}
where $p_{v}$ is the eigenvalue of the operator $-i\partial_{v}$ the
explicit form of the spinor $\eta$ can be obtained after noticing that
the
eigenvalue $K$ corresponds to the operator $\hat{K}_{2}$ given by
equation (%
\ref{31}). Then we have,
\begin{equation}
\label{sol}\eta=\pmatrix{\eta_{1}\cr \eta_{2}\cr}, \qquad
\left(-\sigma^{1}\partial_{y}+\frac{\sigma^{2}}{c}\partial_{z}\right)\eta

=-iK\eta, \qquad \eta_{2}\propto \sigma^{3}\eta_{1}
\end{equation}
substituting the explicit form of the Pauli matrices into (\ref{sol}) we
have
\begin{equation}
\label{prim}\left(\partial_{y}-\frac{k_{z}}{c}\right)\eta_{2}=
iK\eta_{1}
\end{equation}
\begin{equation}
\label{seg}\left(\partial_{y}+\frac{k_{z}}{c}\right)\eta_{1}= iK\eta_{2}

\end{equation}
and for $c(y)=y$ $\eta_{1}$ and $\eta_{2}$ take the form
\begin{equation}
\eta_{1}=y^{1/2}Z_{k_{z}+1/2}(Ky), \qquad
\eta_{2}=-iy^{1/2}Z_{k_{z}-1/2}(Ky)
\end{equation}

\section{Concluding Remarks}

The results presented in this paper can be summarize as follow: applying
the
algebraic method of separation, it is possible to achieve a complete
separation of variables for the Dirac in the background field given by
the
metric (\ref{1}). The system of ordinary differential equations obtained
after the complete separation can be solved in terms of special
functions
for some homogeneous space-times, and in particular for the de-Sitter
and
anti de-Sitter universes. The exact solutions computed in the article
encourage us to analyze more complicate configurations where rotation
would
be present. Also it would interesting to study the problem of separation
for
coupled Einstein-Maxwell sources. Finally, it should be mentioned that
the
exact solutions exhibited in this article could be of help in analyzing
quantum effects associated with spin 1/2 particles in de Sitter space
\cite
{Mishima,Nakayama}.

\end{document}